# ALIGNMENT AND APERTURE SCANS AT THE FERMILAB BOOSTER*

K. Seiya[#], J. Lackey, W Marsh, W. Pellico, , D. Still, K. Triplet, A. Waller

Fermilab, Batavia, IL 60510, USA

*Abstract*

The Fermilab Booster is currently in the process of an intensity upgrade referred to as the Proton Improvement Plan (PIP)[1]. The goal of PIP is to have the Booster provide a proton beam flux of $2\times10^{17}$ protons/hour. This is almost double the current operation of $1.1\times10^{17}$ protons/hour. Beam losses in the machine due to the increased flux will create larger integrated doses on aperture limiting components that will need to be mitigated. The Booster accelerates beam from 400 MeV to 8 GeV at a rep rate of 15hz and then extracts beam to the Main Injector. Several percent of the beam is lost within 3 msec after injection in the early part of acceleration. The aperture at injection energy was recently measured using corrector scans. Along with magnet survey data and aperture scan data a plan to realign the magnets in the Booster was developed and implemented in May 2012. The beam studies, analysis of the scan and alignment data, and the result of the magnet moves are presented.

## PIP INTENSITY UP GRADE

The Fermilab Booster is scheduled to increase the beam cycle rate from 7.7 to 15Hz by 2014 for the future physics program. The Booster is a synchrotron and ramped with a resonance circuit at 15Hz[2]. Protons intensities of $5\times10^{12}$ particles per pulse are accelerated from 400MeV to 8GeV with a cycle efficiency of 90%. Most of the beam loss occurs within 3 msec after injection in the early part of the accelerating ramp. One of the goals of the PIP is to increase the aperture at injection by realigning magnets using survey data and aperture scan results.

## BOOSTER LATTICE AND ALIGNMENT

The Booster has 96 combined function magnets that are located in 24 periods. One period is made up of a 6 m long straight section, a defocusing magnet (D magnet), a 0.5 m mini straight section, a focusing magnet ( F magnet), a 1.2 m short straight section, an F magnet, mini straight section and a D magnet. A combined function magnets has a length of 2.9m and a physical vertical gap of 41.66 mm in D magnets and 57.17 mm in F magnets respectively as shown in Figure 1. There are forty eight corrector magnets total, one located at every end of a long and short straight section. Additional accelerator components such as kickers, collimators and RF are mainly located at the long straights sections (Figure 2). Typical components that can contribute to limiting aperture and their sizes are listed in Table1. Since the size of beam pipe is same as the vertical gap of the D magnet, the compartments have to be perfectly aligned in order to maximize the physical aperture.

Table 1: Typical component aperture size

| Location | Component | Size (H mm * V mm) |
|---|---|---|
| L5 | Pinger, Notcher | 57.17 (circle) |
| L6-7 | Collimator | 76.2* 76.2 |
| L12 | Ext. kicker | 63.25 * 56.85 |
| L14-L23 | RF cavity | 57.17 (circle) |
| L&S1-24 | Corrector | 114.3 (circle) |

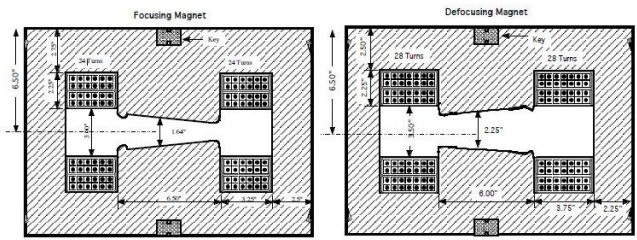

Figure 1: Profile of an F (left) and a D (right) Booster combined function magnet.

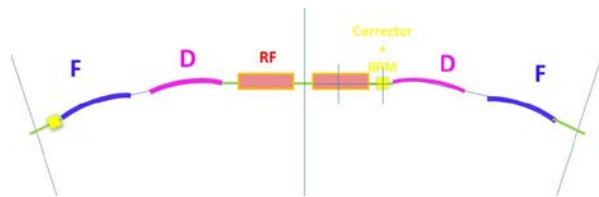

Figure 2: Layout of one Booster period. Beam goes through from left to the right.

Figure 3 shows the lattice function at one period in the Booster. Typical horizontal and vertical beam sizes are $14\pi$ mm-mrad and $16\pi$ mm-mrad respectively. Maximum horizontal beam size is +/-21.5 mm at the short straight sections and the maximum vertical beam size is +/-17.9 mm at long straight sections.

*Survey data*

Alignment data from August 2005 was analyzed for each magnet and the horizontal error is seen in Figure 4.

___________________________________________
* Work supported by Fermi Research Alliance, LLC, under contract No. DE-AC02-07CH11359 with the U.S. Department of Energy.

The vertical error is seen in Figure 5. The F magnets are located inside of the design by 5 mm and the D magnets are located outside of the design by 5 mm in horizontal.

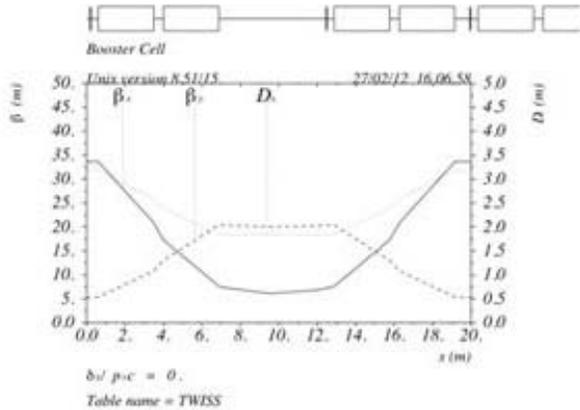

Figure 3: Lattice function at one period in the Booster.

The variation of the error is within +/- 3 mm and shows that both magnets form an elliptical shape instead of a circle. The error in vertical plane shows a 3-5 mm step at few locations. Since the vertical aperture is tight compared to the beam size at long straight section, magnet realignment in vertical should reduce beam loss.

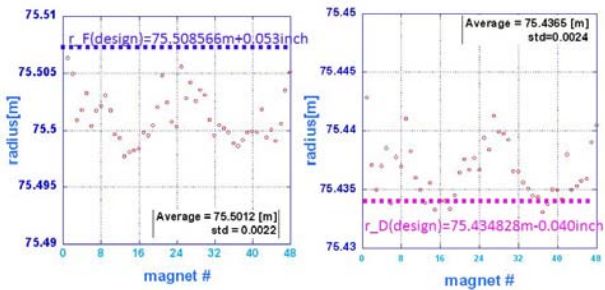

Figure 4: Radial position of the center of the F magnets(left) and D magnets(right).

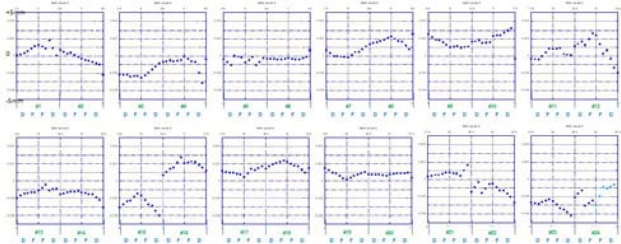

Figure 5: Elevation of the magnet at beginning edge, center and ending edge. The data in one horizontal division has two magnets and show pitch on each magnet.

## APERTURE SCAN

Aperture scans were performed at low intensity using a combination of dipole corrector 3 and 5 bumps. An existing aperture program was modified for the new correctors installed in 2009. The intensity was measured at injection and extraction and compared with the one at the end of the LINAC. An intensity ratio between before and after injection was plotted with the color scales and plotted as shown in Figure 6 (left). Beam position data from beam position monitors (BPM) which are mounted under every corrector was measured at the same time. Since the beam was lost and there was no data from BPM at the edge of the aperture, the corrector current was converted to beam position using a linear response. The data was corrected offline and plotted in Figure 6(right).

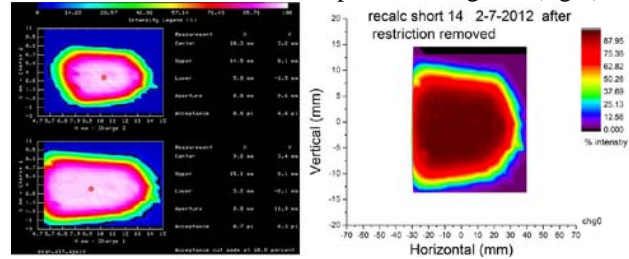

Figure 6: Aperture scan results: %intensity for injection and extraction (left). Corrected for BPM offset. (right).

The aperture was measured at all long and short locations from September 2011 to March 2012. A profile from the scan was fit to the Gaussian distribution and estimated a 90% of the aperture in both horizontal and vertical. Measurements were incomplete at period 24 to period 3 because of the injection bump at L1. Figure 7 (left) and (right) show the horizontal and vertical aperture from the measurements. The horizontal aperture was measured to be -25/+30 mm at short sections and -15/+15 mm at long sections. There are known limiting apertures at injection l at periods 1 to 3 and collimator locations at period 6 and 7. The vertical aperture was measured to be -7/+7 mm at these short sections and -7/+12 mm at long sections. This is smaller than the beam size. The beam orbit was moved to the center of the aperture using the booster orbit smoothing program. Moving the beam to the aperture center increased the intensity by 1%, but increased beam losses at RF cavity locations. High intensity beam studies and collimator retuning will be required in the future.

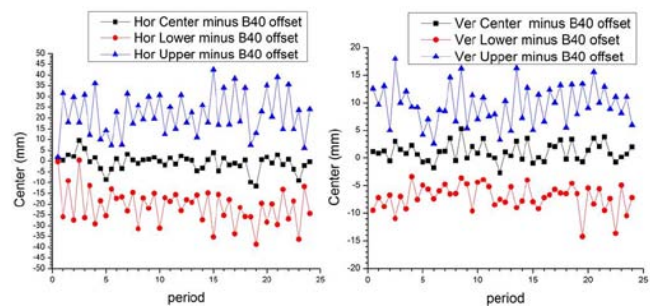

Figure 7: Horizontal (left) and vertical (right) apertures from the measurements: Blue: 90% of the upper aperture, Red: 90% of the lower aperture, Black: center of the aperture

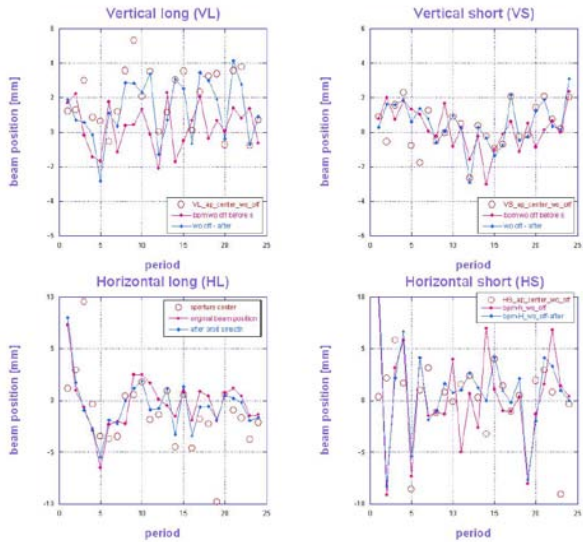

Figure 8: Center of the aperture (open circle), beam positions before(magenta) and after(blue) orbit was moved to the center of the aperture.

## MAGNET REALIGMENT

Two magnets were realigned in May 2012 prior to a long shutdown. The downstream edge of the downstream D magnet at period 21 was lowered by 1.3 mm and the upstream edge of the D magnet at period 22 was raised by 1.3 mm.

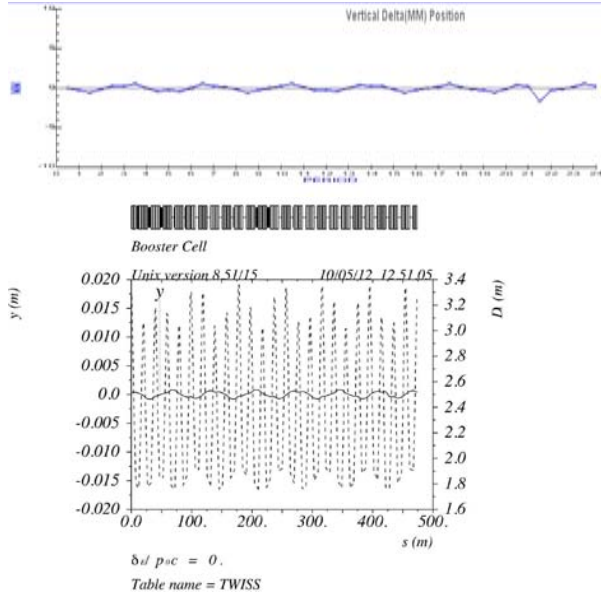

Figure 9: Simulation results of the beam orbit difference with the vertical alignment error Magnet Move Program (upper) and MAD(lower).

The beam orbit with the vertical alignment error was simulated using two different codes, MAD[3] and Magnet Move Program[4] to verify the magnet move. Both results showed within +/-1 mm orbit difference between with and without magnet move. The orbit was measured before and after the magnet move and the differential orbit was +/-1 mm as shown in Figure 10. Figure 11 shows the aperture was increased by ~3 mm as expected after the magnet move.

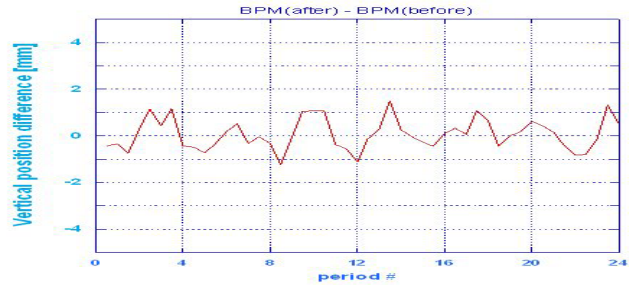

Figure 10: Difference of the measured orbit between before and after a magnet move.

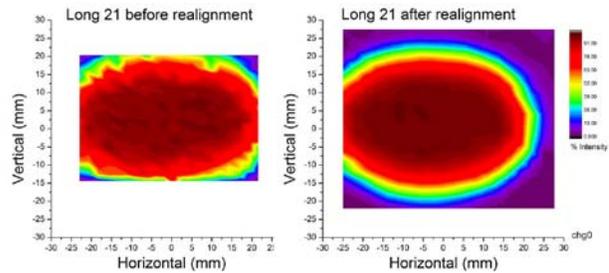

Figure 11: Measured aperture before (left) and after(right) the magnet realignment.

## SUMMARY

The Fermilab Booster has increased the aperture at injection by realigning magnets in preparation for the planned higher intensity operation. Aperture scans were performed at low intensity using standard corrector 3 and 5 bumps. An existing aperture program has been modified to be used with the newly installed (2009) ramped correctors. Alignment data from August 2005 was analyzed and compared with the apertures. The aperture was increased by ~3 mm as expected by centering the beam in the aperture and magnet realignment.